# Towards Product Lining Model-Driven Development Code Generators

Alexander Roth and Bernhard Rumpe

*Software Engineering, RWTH Aachen University, Aachen, Germany*
{*roth,rumpe*}*@se-rwth.de*

Abstract: A code generator systematically transforms compact models to detailed code. Today, code generation is regarded as an integral part of model-driven development (MDD). Despite its relevance, the development of code generators is an inherently complex task and common methodologies and architectures are lacking. Additionally, reuse and extension of existing code generators only exist on individual parts. A systematic development and reuse based on a code generator product line is still in its infancy. Thus, the aim of this paper is to identify the mechanism necessary for a code generator product line by (a) analyzing the common product line development approach and (b) mapping those to a code generator specific infrastructure. As a first step towards realizing a code generator product line infrastructure, we present a component-based implementation approach based on ideas of variability-aware module systems and point out further research challenges.

## 1 INTRODUCTION

An integral part of model-driven development (MDD) are generators, which successively, systematically, and automatically transform abstract models to concrete code. Today's code generators help to "compute an efficient implementation for a nice and easy-to-understand specification" [Czarnecki and Eisenecker, 2000]. Disparate domains have adopted code generators including cyber-physical systems [Haber et al., 2012] and cloud-based systems [Navarro Pérez and Rumpe, 2013].

Even though code generators are used in multiple domains, they are tailored to one domain, serve a purpose, and are mostly not designed for reuse. Product line engineering focuses on identifying commonalities and variability (the ability to change or customize a system in a predefined way [Pohl et al., 2005]) to create components that are used to develop software products for an area of application [Clements and Northrop, 2012]. Applying ideas of product line engineering to code generators helps to design reusable code generator components in order to develop generator product lines and support reuse.

Multiple authors have discussed implementation mechanisms for code generator product lines [Batory et al., 1994, Czarnecki and Eisenecker, 2000, Smaragdakis and Batory, 2000, Völter and Groher, 2007a, Kulkarni and Reddy, 2008, Jörges, 2013]. However, the approaches either require the developer to know the code generator's internals [Apel et al., 2013]; result in hardly maintainable code generators; are lacking essential concepts to prevent dublicate components with minor adaptations; or are based on textual composition of generated artifacts. These short comings in code generator product line development result in hardly maintainable and understandable code generators and hamper code generator reuse.

Consequently, the aim of this paper is to apply current ideas of product line engineering to MDD code generators in order to identify research opportunities. By analyzing common approaches to implement product lines a mapping to a code generator infrastructure is presented. This infrastructure is based on code generator components with a separation of global and local variability [Kästner et al., 2012]. We argue that a clear understanding and a language independent implementation concept of a code generator component's interface is crucial for code generator product lines and present possible extensions.

Hence, an overview of current literature (Section 2) shows the state of the art and points out short comings in this research area. Next, our understanding of a code generator is presented in Section 3 to discuss upcoming challenges in developing code generator product lines. By applying variability-aware module systems to code generators further challenges including code generator component interfaces are presented in Section 4. Finally, code generator component composition with respect to the differ-ent forms of composition is introduced in Section 5 and the paper is concluded in Section 6.

## 2 RELATED WORK

Preprocessor and conditional approaches are the most popular approaches to implement variability [Czarnecki and Eisenecker, 2000, Völter and Groher, 2007a]. To avoid hardly maintainable code because of too many conditions [Spencer and Collyer, 1992], multiple approaches have been proposed including an approach based on Aspect-Oriented Programming [Völter and Groher, 2007a, Völter and Groher, 2007b], transformation-based approaches [Smaragdakis and Batory, 2000], and domain specific language-based approaches [Singhal, 1996, Czarnecki et al., 2000]. A major drawback of these approaches is that in depth knowledge of the generator's internals is required [Apel et al., 2013]. Moreover, the language based approaches require either a powerful transformation or domain-specific language. Rather than proposing a language to implement variability, this paper focuses on identifying research directions in developing code generator product lines and proposes explicit interfaces to support information hiding.

Besides the above variability implementation mechanisms that have been used to implement code generators, a component-based [Kulkarni and Reddy, 2008], a layered [Batory and O'Malley, 1992], and a service-oriented approach [Jörges, 2013] have been proposed to implement code generator product lines. These approaches are based on textual composition of artifacts. Additionally, in the latter approach a variant is a full copy of an existing code generator with additional extensions and adaptations. Thereby, the commonalities of all variants are not managed on their. Moreover, none of the approaches addresses input language variability, which is the variability defined by the input model, e.g. by defining a keyword to change the generated code. In this paper, we build on top of these ideas but consider input language variability as well as variability within components.

## 3 GENERATORS AND PRODUCT LINES

A *generator* can be seen as a software system that produces an implementation from a higher-level description of a part of a software [Czarnecki and Eisenecker, 2000]. A *code generator* is a special type of a generator that creates an implementation in a programming language from a set of input artifacts, which are typically models. We also assume that a code generator always terminates, is deterministic, is not an interactive system, and generates at least one output artifact from one input artifact. Such code generators are built on top of existing compilers for programming languages [Brambilla et al., 2012].

Basically, a code generator consists of a front-end (*language processing*) and a back-end (*code generation*). Language processing is concerned with parsing the models, checking language constraints and creating an internal representation (*abstract syntax tree* and *symboltable*). Code generation transforms the internal representation to concrete code stored in output artifacts.

In most cases, a code generator has one particular purpose and is not designed in an extensible way for being reused it in different application areas. An approach to exploit code generator reuse, i.e., allowing customizations and adaptations to the code generator to redefine the purpose or the area of application, is to create a code generator product line. In the remainder of this section, we introduce code generator product lines and identify research opportunities.

### 3.1 Code Generator Product Lines

A *code generator product line* is a set of components used to create a concrete code generator product that describes a software product line for a specific area of application. A *code generator variant*, which represents a concrete code generator, is modeled by selecting a set of components and adding customizations and adaptations. A code generator product line strongly differs from general product line development as understood in [Clements and Northrop, 2012, Bosch, 2000, Pohl et al., 2005], because (a) code generators are techniques to manage end-product lines and (b) a code generator is often at least partially itself generated. Based on these primary differences code generator product lines can be seen as either a tooling to configure software product lines or as a product line on its own. In the remainder of this paper, we use the latter understanding.

Each code generator product line component is either a front-end or a back-end component. Front-end components are concerned with extending the language processing, e.g. additional constraints to restrict the set of input models. A back-end component adds functionality to the code generation process.

In the following, we shortly describe the two processes in which a product line is developed and point out open research questions.

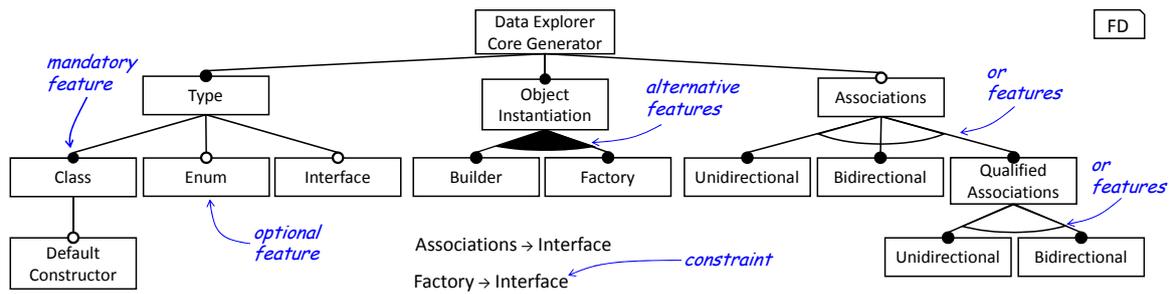

Figure 1: A a feature model of a simplified class diagram-to-Java code generator.

### 3.1.1 Domain Engineering

Domain engineering focuses on collecting, organizing, and storing past experience in developing systems in a domain in the form of reusable assets and additionally providing means of reusing these assets [Czarnecki and Eisenecker, 2000]. All relevant information is integrated in a coherent domain model. An example of the commonalities and variability of a code generator product line for a class diagram-to-Java generator is shown in the feature model [Kang et al., 1990] in Figure 1.

A research question is if domain engineering for code generator product lines differs from domain engineering proposed for software product lines. For instance, since each code generator maps concepts of the input language to concepts of the target language features of the input language and features of the output language are part of the code generator feature model. In Figure 1 input language concepts (Class, Enum, and Interface features) as well as output language concepts (Builder and Factory) are part of the code generator product line.

In the domain engineering the architecture of the product line is designed as well. To our knowledge GenVoca [Batory et al., 1994], a component-based approach [Kulkarni and Reddy, 2008], and a Service-Oriented architecture [Jörges, 2013] are currently the only architecture proposed for generators. Possible candidates for other code generator product line architectures are generic architectures that define a fixed frame with a number of extension points and highly flexible architectures that support structural variation, i.e., even the skeleton may vary.

By analyzing existing approaches to implement software product line architectures, we have extracted a set of requirements that should be considered when developing a code generator product line:

1. *Traceability of Features.* Throughout the lifecycle of the product line a mapping between a feature and its implementing source code elements should be kept to ensure maintainability [Apel et al., 2013].

2. *Selection and Configuration of Features.* A selection and configuration mechanism has to be easy to understand, has to detect invalid selections and configurations, and has to directly influence the resulting product, i.e., every selected feature must be included in the product.

3. *Separation of Concerns.* The implementation of a product line should be separated into different components to support component reusability [Clements and Northrop, 2012].

4. *Separation of Variability.* Separating variability by allowing each component to be a product line on its own helps reducing complexity [Kästner et al., 2012]. In generator product lines variability at generator level and variability at generated artifact level have to be regarded as well.

5. *Enable Reuse.* Product line development aims at designing components that can be reused in other product lines [Pohl et al., 2005, Clements and Northrop, 2012].

6. *Component Composition.* A product line implementation regards component composition by making it "as easy as merging code" [Kästner et al., 2012]. From set of components a variant is composed by first validating constraints and second sticking the components together.

Subsequently, we analyze the process of how to form a code generator variant and point out research opportunities.

### 3.1.2 Application Engineering

In the application engineering process a variant is derived using the product line architecture and the set of components. After selecting and configuring the product line components, customizations are added to define a concrete product variant. In code generator product lines customizing components means to adapt the components' behavior by including handwritten extensions or writing glue code. An infrastructure for code generator product lines has to support such kind of extensions.

The derivation process of a variant can either be performed manually, automated (assembly of components is supported by tools and generators), or automatically (ordering and assembly of the variant is supported by tools and generators [Czarnecki and Eisenecker, 2000]). Here, the automatical approach extends the automated approach by deriving a variant from a specification. In all of these approaches, a general challenge is to ensure the syntax, type, and behavioral correctness.

In the remainder, we implement a code generator product line to identify further research challenges.

# 4 IMPLEMENTING CODE GENERATORS PRODUCT LINES

Using the variability-aware module system approach [Kästner et al., 2012], we implement a code generator product line. We are aware of other approaches to implement code generator product lines. This approach has been chosen because it fits the idea of reusable components with separated variability concerns as described in Section 3.1.1.

A variability-aware module system eliminates predominant variability, i.e., the need to represent variability using components, in component-based approaches by separating variability into *global* and *local* variability concerns. Global variability is concerned with representing variability by components, whereas local variability is concerned with the variability inside a component. Each component provides configuration options to configure local variability. Since components contain local variability, they form product lines on their own. Composing components becomes composition of entire product lines.

To give an example of a code generator product line, we assume the feature model in Figure 1 shows our code generator product line. It produced Java code from class diagram models. Rather than using a feature-to-component mapping, we map the feature `Types` to one component with a configuration option for the `Default Constructor`. The component's interface contains *variability concerns* (generate classes, interfaces, enumerations, or default constructors) and *variability constraints* (classes are mandatory). Here, global variability is modeled by the `Types` component and local variability is modeled by the configuration options. A research question is if there are guidelines on how to map features to components with local variability.

Crucial to component-based approaches are interfaces. In the following, we introduce code generator component interfaces.

## 4.1 Code Generator Component Interface

A *code generator component interface* is the interface provided by each code generator component. Besides configuration options for local variability, it may also contain variation points, which are define by developers and assigned during application engineering to specify individual adaptations. These variation points can be used in the application engineering process to customize a component.

For code generator components, it is not practical to define each generated method because (a) they are possibly generated during run-time and, thus, do not exist a priori and (b) rather than method declarations parts of generated code are exchanged to parameterize a code generator component at generation-time. Without explicitly method or code declarations other ways have to be found to prevent generation of syntactically incorrect code. A research direction based on generating code into containers and checking its syntax before writing to a file has been proposed in [Zschaler and Rashid, 2011]. Still it is unknown what methods are generated before generation-time.

During code generation components exchange information to adapt the generation process. For instance, information about generated artifacts are exchanged to prevent overwriting. Typically, such information is implicit and only known to the generator developer. This hampers efficient code generator reuse. To improve reusability such implicit knowledge should be made explicit in code generator component interfaces. However, such an approach requires an ontology for the exchanged information and an infrastructure to support information exchange, e.g. shared-memory [Batory and O'Malley, 1992, Jörges, 2013].

As a code generator component may contain transformations, a code generator component may define different behaviors for different configurations. For instance, before generating code some transformations have to be applied. Such behavior may also be configured during application engineering to, e.g. leaf out some transformations. Thus, an infrastructure demands for explicit representation of component's behavior and support for customizing such behavior.

# 5 CODE GENERATOR COMPONENT COMPOSITION

Deriving a concrete product from a set of components means to compose components and add customiza-tions. In compliance to [Rumpe, 2013], we understand code generator component composition as the derivation of a code generator component $C = A \otimes B$ from two code generator components $A$ and $B$ using a *composition operator* $\otimes$. This composition demands for two code generator components $A$ and $B$ with precisely defined interfaces and encapsulated internals. The composition $C = A \otimes B$ binds both code generator components together with respect to their common interface. The the resulting composition has a combined meaning. A definition of the composition operator for the variability-aware module system that also regards global and local variability has been proposed in [Kästner et al., 2012].

Composition is separated into different forms [Rumpe, 2013]. In the following, we analyze these forms and give implications for a code generator product line infrastructure.

## 5.1 Forms of Composition

*Syntactic composition* deals with how the component composition looks like. When composing code generator components, the composition of two code generator components may generate one or multiple artifact(s) that either contain code that is produced by multiple code generator components or code that is generated by only one code generator component. In both cases the generated code needs to be checked for syntactic errors.

Assuming that semantics is understood as in [Harel and Rumpe, 2004], then *semantic composition* is concerned with the meaning of the code generator component composition. Each code generator component defines the mapping of the input language concepts to concepts in the output language. The meaning of a code generator composition is the composed meaning of all involved code generator components. The semantics are derived from the composition operator and the semantics of each composed code generator component. An infrastructure for code generator product lines has to support individual definitions of the composition operator, since the semantics may vary for different code generator variants.

A major challenge in code generator product line development is to decompose the overall code generator into small components. In consequence, that the code generator component interfaces have to be designed for composition. This form of composition - called *methodical composition* - points out that a method for mapping feature models to code generator components is currently lacking.

When the overall code generator product line is decomposed in code generator components the development task is to be organized. *Organizational composition* is concerned with organizing the collaborative and independent development of code generator components. A prerequisite for an independent development is a clear definition of the code generator component interfaces. This forms a specification that has to be followed by each development team of a code generator component. An initial approach has been proposed in [Kulkarni and Reddy, 2008].

Finally, *technical composition* regards the demands for the infrastructure to enable code generator component composition. A major challenge is incremental code generation, i.e., small changes to the input do not require a complete regeneration but only the relevant parts are regenerated. Technical composition also regards binding times of the code generator components. Conceptually, the components need to be composed as soon as possible in the development such that developers have an understanding of the final product. Technically the composition should be handled as late as possible to support partial regeneration and increase speed of code generation.

In the following, we present different ways to compose code generator components.

## 5.2 Binding Times

Composing code generator components at generation time requires an infrastructure that allows components to exchange information about the generated artifacts or the component's internals, e.g. names of generated methods or pieces of target code. Then, the generated artifacts are created collaboratively. Despite early error detection in the generation process, a common understanding and a standardization of the information exchanged is crucial. Only then, collaboration between code generator components can take place. Moreover, in such an approach information exchange may also require knowledge about the context of the exchanged information.

Another approach to compose code generator components is to create artifacts that can be composed at run-time. The generated artifacts are a priori designed to provide target language constructs that are used in other artifacts. For instance, provide predefined methods to call functionality. This approach suffers from implicit knowledge about the generated artifacts. It is only beneficial for more than one developer, if the implicit knowledge is available for every developer at anytime.

Finally, a hybrid approach that uses composition at generation- and run-time may provide a sufficient solution to handle composition. However, it is currently not common in practice.

# 6 CONCLUSION

Despite the relevance of MDD code generators, there are no common development methodologies and approaches for developing code generators product lines. We have analyzed the common processes of product line engineering to identify research challenges for code generator product lines. It shows that research is needed in analyzing, designing architectures, and implementing code generator product lines.

As a primary step towards an infrastructure for code generator product lines, we have applied ideas of variability-aware module systems to identify research challenges. Crucial to a code generator product line are interfaces, which contain variability, generation information, extensions to the generated artifacts, and different behavior of the code generator components. This is unique to code generator product lines compared to general software product lines.

We have pointed out the different forms of code generator composition to derive requirements for a code generator product line infrastructure. The main requirements for a code generator product line infrastructure are support for incremental code generation, specification of code generator component interfaces, support for validation of generated code, and support for individual semantics of a composition operator.